\documentclass[showpacs,APS]{revtex4}

\usepackage{graphicx}
\usepackage{amsmath}
\usepackage{color}
\usepackage[normalem]{ulem}
\usepackage{multirow}

\DeclareRobustCommand\bblash{\btt{\@backslashchar}} \makeatother

\begin{document}
\title{On the equilibrium of the Buchdahl star }
\author{Naresh Dadhich\footnote{nkd@iucaa.in}}
\affiliation{Inter-University Centre for Astronomy \& Astrophysics,\\ Post Bag 4, Pune, 411 007, India}
\date{\today}

\begin{abstract}
The Buchdahl star is the limiting compactness (which is indicated by sturation of the Buchdahl bound) object without horizon. It is in general defined by the potential felt by radially falling timelike particle,  $\Phi(R) = 4/9$, in the field of a static object.  On the other hand black hole is similarly characterized by  $\Phi(R)=1/2$ which defines  the horizon. Further it is remarkable that in terms of gravitational and non-gravitational energy, the Buchdahl star is alternatively defined when gravitational energy is half of non-gravitational energy while the black hole when the two are equal. When an infinitely dispersed system of bare mass $M$ collapses under its own gravity to radius $R$, total energy encompassed inside $R$ would be $E_{tot}(R)=M-E_{grav}(R)$. That is, energy inside the object is increased by the amount equivalent to gravitational energy lying outside and which manifests  as internal energy in the interior. If the interior consists of free particles in motion interacting only through gravity as is the case for the Vlasov kinetic matter, internal (gravitational) energy could be thought of as kinetic energy and the defining condition for the  Buchdahl star would then be kinetic (gravitational) energy equal to half of non-gravitational (potential) energy. Consequently it could be envisaged that equilibrium of the Buchdahl star interior is governed by the celebrated Virial theorem like relation (average kinetic energy equal to half of average potential energy). On the same count the black hole equilibrium is governed by equality of gravitational and non-gravitational energy ! 
\end{abstract}

\pacs {04.07, 04.70 Bw, 97.60.Lf}

\maketitle

\section{Introduction}\

With the unprecedented pathbreaking and precision advances in observations in high energy astrophysics and cosmology following the monumental discovery of gravitational waves \cite{gw-dis} produced by merger of two black holes, and almost hearing the ringing down of black hole through quasi-normal modes has heralded the advent of gravitational wave astronomy with a bang. The new discoveries have been showering in at regular intervals in terms of black hole - black hole, black hole - neutron star as well as neutron star - neutron star mergers. Then there are the fantastic observations of an object as compact as black hole \cite{bh-dis} and of black hole shadow \cite{bh-sha}. These are the most glorious and remarkable discoveries of all times. The high energy observational astrophysics involving the compact objects like black holes and neutron stars has come into full bloom. With this backdrop it is natural to ask how compact a star like object in principle could be? The obvious answer in general relativity (GR) is black hole (BH) with a horizon which unfortunately blocks out all the information of its interior. To have some idea and understand physics very close to horizon, one would like to have some non-null construct, very close to BH which could be in interaction and communication with outside. That is why the notions of membrane paradigm and stretched horizon were invoked \cite{thorne} in which a non-null surface very close to horizon, but not a horizon, is envisaged for which black hole  physics and thermodynamics could be explored. That is to have a timelike surface as close as possible to horizon which could be in active physical communication with outside world. \\

There could in fact exist the most compact  astrophysical object without horizon in the Buchdahl star (BS), defined in general by $\Phi(R)=4/9$ where $Phi(R)$ is the potential felt by a radially falling particle \cite{maxforce}. It is identified with the equality in the Buchdahl compactness bound $M/R \leq 4/9$ \cite{buch} which is in general given by $Phi(R) \leq 4/9$ for a static object whether charged or neutral. This bound was derived by Buchdahl \cite{buch} under very general conditions of isotropic perfect fluid with density decreasing outwards and fluid interior at the boundary is matched to the exterior Schwarzschild vacuum. For inclusion of charge $M/R$ is replaced by $\Phi(R) = (M - Q^2/2R)/R$. Similarly black hole (BH) is defined by  $\Phi(R) = 1/2$.   It should be noted that the metric in the static case describes field of a static object irrespective of it being BH or not and hence it is equally valid for BS. This is however not the case for the axially symmetric Kerr metric which could only describe a rotating BH and not a non-BH rotating object like a rotating BS. It may however be noted that we could similarly compute the potential for axial fall of timelike particle, which would read as $\Phi(R) = M/R(1 + a^2/R^2)$ where $a$ indicates spin of the object. Then $\Phi(R) = 1/2, 4/9$ would give a rotating BH and BS. This is all fine for BH but not for BS because unlike the spherically symmetric case, the axially symmetric Kerr metric cannot describe a non-BH rotating object. This is because rotation would give rise to flattening of the object at the poles and consequently to multipole moments which the Kerr metric cannot harbour. For BH, all the moments are radiated away before null horizon is formed and only mass, charge and spin survive -- the well known No Hair theorem \cite{MTW}.  Since BS has timlike boundary and hence it has to have multipole moments which cannot be sustained by  the Kerr metric. There however exits no exact solution for description of a rotating object with multipole moments except the one due to Hartle and Thorne \cite{har-tho} in the slow rotation approximation. We would therefore stick to the static case only leaving out the rotating case from any further consideration.   \\

Note that the Buchdahl bound is also given by pressure at the centre $p_c(R=0)\leq\infty$ for the stiffest equation of state of uniform density --- incompressible perfect fluid distribution described by the unique Schwarzschild's interior solution not only for the Einstein but also for the Lovelock gravity in general \cite{dmk10}. The same limit was also obtained \cite{andrea08, kara-stal08} by assuming the strong energy condition, $p_r + 2p_t \leq \rho$, where $p_r, \, p_t, \, \rho$ are respectively radial and transverse pressure, and matter energy density. In this case the limit saturates not for fluid interior but for an infinitely thin shell for which all the three quantities diverge at the boundary and in the limit $p_r/\rho \to 0$ and $2p_t/\rho \to 1$ \cite{andrea08}. In the previous case the limit saturated for central pressure tending to infinity while in the latter, it is for infinitely thin shell. In either case, it is a limiting case. \\

There exists extensive literature on alternative derivations of the bound involving various situations like inclusion of $\Lambda$ \cite{mak-stuch}, different conditions than Buchdahl's \cite{andrea09, kara-stal08}, brane-world gravity \cite{ger-maa, gar-ure}, modified gravity theories including Lovelock gravity and higher dimensions \cite{gos, wri, sum-dad16, Feng18}. Further there are several works finding compactness bound for a charged star \cite{roth, mak-dob-har,  boe-har, kara-stal08, andrea09, and-ekl-rei, lemos14, coop, florides, guil, lemos10, lemos14a}. Unlike the neutral case, here they all do not agree. However there is one \cite{roth} that accord to the general universal prescription of $\Phi(R) = (M- Q^2/2R)/R \leq 4/9$ and it reads as follows:
\begin{equation}
M/R \leq \frac{8/9}{1 + \sqrt{1 - (8/9)\alpha^2}}~, \, \alpha^2 = Q^2/M^2~.
\end{equation}
It reduces to the Buchdahl bound for the uncharged case, $M/R \leq 4/9$ for $Q = 0 $. Like BH, it also has extremal limit, $\alpha^2 = 9/8 >1$ while $M/R \leq 8/9 < 1$ always . This means BS could indeed have  $\alpha^2 >1$; i.e. overcharged relative to BH. However its radius is always greater than the BH horizon. The bound could be further decreased by appealing to the dominant energy condition and sound velocity being subliminal \cite{bar}. The Buchdahl bound defines an overriding state which is obtained under very general conditions while more compact distributions are allowed under specific circumstances and conditions. However it has recently been rigorously established \cite{alho22} that the Buchdahl bound cannot be pierced for a fluid obeying the energy conditions and radial stability of the distribution as well as the sound velocity remaining subluminal. Thus the Buchdahl bound has to be always respected by any physically reasonable matter field interior for a compact star. The Buchdahl star is the limiting case when this bound saturates, and it is thus the most compact object without horizon. \\

Before we go any further we should however mention that BH has a geometric definition in the timelike Killing vector becoming null defining the event horizon. It is the null character of the vector where the norm of vector equal to zero that yields the fixed value of radius for BH horizon. In contrast BS has a timelike boundary surface defined by its normal being timelike (norm being $>1$) which cannot give a fixed value of the boundary radius. We therefore resort to the simple physical property that velocity of an infalling particle which is given by $v^2 = 2\Phi(R)$ attains the value unity at the BH horizon which is in fact equivalent to the timelike Killing vector turning null. On the other hand the BS boundary radius is given by $v^2=2\Phi(R)=8/9$. This is the simple physical argument for characterizing BH and BS respectively by $\Phi(R)= 1/2, 4/9$. \\ 

Now we come to gravitational energy and its role in the present discussion of the compactness bound. Energy is a very illdefined and ambiguous entity in general relativity, and more so gravitational energy which resides in space curvature and hence is in principle non-localizable and therefore hard to compute unambiguously. However there is the Brown-York prescription \cite{bro-yor} of quasilocal energy that gives a good measure of total energy contained inside a sphere of radius $R$ enclosing a gravitating body. In this prescription, it is envisaged that an infinitely dispersed system of bare mass $M$ begins collapsing under its own gravity and as collapse proceeds it picks up gravitational field energy which lies exterior to the object. Then total energy contained inside would be given by $E_{tot}(R) = M - E_{grav}(R)$, and thus gravitational energy, $E_{grav}$ could be computed. How far collapse goes on and where it halts would determine bound on compactness as well as on gravitational energy. This would mean that the compactness bound could be obtained as a bound on $E_{grav}$, which could be computed by using the exterior metric alone without any reference to what happens inside the object. Further the exterior solution is unique for a static object, in general given by the  Reissner - Nordstr{$\ddot o$}m metric describing the field of a charged object. \\

By using the Brown-York prescription \cite{bro-yor} of quasilocal energy for computing $E_{grav}$, we had long back \cite{dad97} defined BH horizon by equipartition of total energy into gravitational and non-gravitational energy. That is, horizon is defined when  gravitational energy equals non-gravitational energy. This could easily be understood as follows: at horizon timelike particle that feels gravitational attraction produced by  $E_{non-grav}$ tends to be photon which could only feel space curvature produced by $E_{grav}$ \cite{dad12}. That is why at the horizon, where timelike particle tends to null, the two sources they respond to should also have the same measure; i.e., $E_{grav} = E_{non-grav}$. The equipartition of total energy into gravitational and non-gravitational thus defines the BH horizon in general. This was a remarkably insightful and physically illuminating explanation and understanding of BH horizon anchored on gravitational and non-gravitational energy. \\

The question then arises, could we similarly appeal to some general principle or property for obtaining the compactness bound for a non-BH object? The increase in energy equivalent to $E_{grav}$ in the interior manifests as internal energy which counters gravity produced by $E_{non-grav}$, and a balance between the two would define the equilibrium. The Buchdahl compactness bound and BS would be determined by the ratio, $E_{grav}/E_{non-grav} = \gamma$ attaining  certain critical value. For BH, $\gamma =1$, what should it be for non-BH object like BS ? First thing that comes to mind is the Virial theorem expressing the equilibrium by the ratio, kinetic to potential energy being half for a cloud of free particles interacting only through gravity. For this kinetic distribution, internal energy which is equal to $E_{grav}$ would be kinetic energy and $E_{non-grav}$ would be analogous to potential energy. Then $\gamma=1/2$ indeed \cite{dad19}, as we show in the following, gives the saturated Buchdahl bound $\Phi(R)=4/9$ defining the Buchdhal star. It is remarkable that its equilibrium is governed by the well known Virial theorem like relation. That is, as BH is defined when gravitational energy is equal to non-gravitational energy, similarly BS is defined when the former is half of the latter. This is then equally insightful and illuminating. Aboveall all this is computed by employing the unique exterior metric without reference to the interior distribution. This is also remarkable that the compactness bound could be computed from the unique extrior metric with no reference to interior distribution at all. The main aim of the paper is the realization the critical role of gravitational energy in obtaining the Buchdahl bound and thereby BS \cite{dad19} and its equilibrium being governed by the celebrated Virial theorem. \\

In what follows, in the next Sec. we shall briefly recall the Brown-York quasilocal energy and by using that we compute gravitational energy. It then turns out that the BS defining condition $\Phi(R)=4/9$ is equivalent to gravitational energy being half of non-gravitational energy. This clearly envisages the Virial equilibrium condition which is then taken up next followed by the discussion. \\

\section{Buchdahl star from gravitational energy} 

We shall consider the static object having mass $M$ and charge $Q$ described by the Reissner - Nordstr{$\ddot o$}m metric, $g_{tt} = -1/g_{rr} = 1 - 2\Phi(R) = 1 - 2M/R + Q^2/R^2$ where $\Phi(R) = (M - Q^2/2R)/R$. This clearly indicates the electric field energy lying extrior to $R$ has to be subtracted out from $M$ to write the potential. All that follows would be referred to this spacetime metric. \\

\subsection{Brown-York quasilocal energy}

Let us briefly recall the Brown-York prescription \cite{bro-yor} in which it is envisioned that a space-time region is bounded in a $3$-cylindrical timelike surface bounded at the two ends by a $2$-surface. Then the Brown-York quasilocal energy is defined by
\begin{equation}
E_{BY}= \frac{1}{8\pi} \int{d^2x\sqrt{q}(k-k_0)} = E_{tot}(R)
\end{equation}
where $k$ and $q$ are respectively trace of extrinsic curvature and determinant of the metric, $q_{ab}$ on $2$-surface. The reference extrinsic curvature, $k_0$ is of some reference space-time, which for asymptotically flat case would naturally be the Minkowski flat. This is the measure of total energy contained inside a sphere of some radius $R$ around a static object.  The evaluation of the above integral for the Reissner - Nordstr{$\ddot o$}m metric yields,
\begin{equation}
E_{tot}(R) = R - \sqrt{R^2 - 2MR + Q^2},
\end{equation}
which at large $R$ approximates to
\begin{equation}
E_{tot}(R) = M - (Q^2/2R - M^2/2R) = M - Q^2/2R + M^2/2R .
\end{equation}
This prescription envisions an infinitely dispersed distribution of bare ADM mass $M$ \cite{ADM} at infinity, while collapsing under its own gravity it picks up gravitational field energy as well as electrostatic field energy. Then gravitational field energy at $R$ would be given by subtracting from it mass $M$ and electrostatic energy $Q^2/2R$ lying outside $R$. That is to evaluate gravitational field energy, subtract non-gravitational matter energy $E_{non-grav} = M - Q^2/2R$ from total energy, $E_{tot}(R)$ contained inside radius $R$. Thus gravitational field energy lying outside $R$ is given by
\begin{equation}
E_{grav}(R) = E_{tot} - (M - Q^2/2R) = R - \sqrt{R^2 - 2MR + Q^2} - (M - Q^2/2R).
\end{equation}
Note that total energy, $E_{tot}(R) = E_{non-grav}(R) + E_{grav}(R)$. \\

The Brown-York prescription of quasilocal energy is good not only because it gives the expected result for gravitational energy in the first approximation but also it has attracted considerable  attention in the literature, and like mass its positivity has also been proven \cite{liu-yau}. \\

Writing the Brown-York quasilocal energy giving total energy contained inside radius $R$ as
\begin{equation}
 E_{tot}(R) = R(1 - \sqrt{1 - 2\Phi(R)})
\end{equation}
where $\Phi(R) = E_{non-grav}/R$ and $E_{non-grav} = M-Q^2/2R$ for charged object. Expanding the above relation for large $R$ for $Q=0$, it is $E_{tot}(R)=M-(-M^2/2R)=M+M^2/2R$ for the neutral static star. This is how gravitational energy gets added to mass $M$ to give total energy contained inside a spherical object. Ultimately it is the balance between gravitational and non-gravitational energy that determines how compact an object could become?. \\ 

\subsection{Buchdahl star} 

Let's write  $E_{grav} = \gamma E_{non-grav}$ which would read as
\begin{equation}
1 - \sqrt{1 - 2\Phi(R)} = (1 + \gamma)\Phi(R)\, ,
\end{equation}
equivalently
\begin{equation}
\Phi(R) = \frac{2\gamma}{(1+\gamma)^2}\, .
\end{equation}

Then the Buchdahl star is given by $\gamma=1/2$ and the bound by $\gamma\leq1/2$; i.e.,  
\begin{equation}
\Phi(R) = \frac{M-Q^2/2R}{R} = \frac{2\gamma}{(1+\gamma)^2} = 4/9\, , 
\end{equation}
and equivalently 
\begin{equation}
M/R = \frac{8/9}{1 + \sqrt{1 - (8/9)\alpha^2}} 
\end{equation}
where $\alpha^2=Q^2/M^2$. It reduces to $M/R=4/9$ for the neutral case. It is BH when $\gamma=1$. \\

The Buchdahl star is characterized by gravitational energy being half of the non-gravitationl energy while the black hole by their equality. 

\section{The Virial equilibrium}

As we have seen above that energy in the interior of an object increases by the amount equivalent to gravitational energy, and that is the measure of its internal (kinetic/binding) energy which may manifest as pressure for fluid distribution or kinetic (thermal) energy (KE) in the special case of the Vlasov kinetic matter or in any other form. It is the balance between gravitational (internal/kinetic) and non-gravitational (potential) energy that governs the equilibrium of star interior. Then the Buchdahl bound condition, $E_{grav} \leq \frac{1}{2}E_{non-grav}$ translates to internal energy being less or equal to half of non-gravitational energy. Buchdahl star is defined when this condition is saturated and its equilibrium is defined by the internal energy being half non-gravitational (matter) energy. In the limiting compactness state, it is conceivable that elastic bonds of fluid may start loosening up giving rise to pure kinetic Vlasov matter consisting of free particles in motion interacting only through gravity. In that case internal energy would indeed be kinetic energy and then its equilibrium would truly be governed by the celebrated Virial theorem, average kinetic energy equal to half average potential energy. It turns out that for this Vlasov kinetic distribution, the Buchdahl bound could only saturate for a thin shell satisfying the  strong energy condition, $p_r/\rho \to 0$ and $2p_t/\rho \to 1$ \cite{andrea08}. Thus we arrive at a very important and remarkable result that BS could indeed be a Virial star with the Vlasov kinetic matter as source and the equilbrium governed by the Virial theorem. \\

This prompts us to envisage that BS interior consists of free particles moving with velocity $V^2= 2Phi(R) = 8/9$ while BH has null particles with $V^2 = 2Phi(R) = 1$. That is, both the limiting compactness objects,  one with hoorizon and the other without, may have free particles in motion (with $v^2=1, 8/9$ respectively for BH and BS) interacting only through gravity confined into a thin shell near the boundary. This seems a plausible and physically appealing scenario which has to be further probed and explored. In this regard, there has been a very interesting attempt  by Mazur and Mottola \cite{maz-mot} for a model of regular BH interior. They take the constant density Schwarzschild solution which has pressure divergence at the centre for BS of radius $R=9/4\,M$ (in particular in the range,  $1/2\geq M/R \geq4/9$), and that this divergence is integrable giving rise to negative constant pressure below certain critical radius. The regular BH interior is then given by the deSitter spacetime with $p=-\rho= const.$.  \\


It may be of interest to find solutions with the linear equation of state $p = k\rho$, and in particular for $k=1/2, 1$. There is a good analysis of this class of solutions in \cite{Tiw17}, and it turns out that distribution cannot be bounded as pressure cannot vanish at any finite radius. On the other hand there is a very simple argument that prohibits the linear equation of state for interior solution. The boundary of interior distribution is identified by $p=0$ which implies vanishing of sound speed there while the equation of state gives  $v_s^2=dp/d\rho=k\neq0$. This could however be easily circumvented by considering the anisotropic distribution with $p_r=0$ and $p_t=k\rho$. It should be interesting to explore such anisotropic models. \\

As we have discussed in the above that for the Virial equilibrium, the most suitable matter field seems to be the Vlasov kinetic matter. There is an extensive investigation of that by Andreasson \cite{And11} and collaborators \cite{And-Rei06, Ame-And-Log19, Ame-And-Log16}. In particular the Vlasov matter interior solution is discussed for a spherical object \cite{And-Rei06}. It turns out that density may not have monotonic behaviour but instead have richer profile having many peaks, as well as being zero at the centre. The stability of these solutions have been examined numerically and it is found that they tend to be unstable as $M/R$ approaches the Buchdahl bound $4/9$ \cite{ Ame-And-Log19, Ame-And-Log16}. These are all numerical results and there is no hard result ruling out existence of stable solutions for $M/R=4/9$. On the other hand axially symmetric solutions with very high rotation may have a greater chance of being stable. In this regard it should be noted that models for highly compact rotating objects have been numerically constructed \cite{ Ame-And-Log19, Ame-And-Log16}. As mentioned earlier the Buchdahl bound stands good and firm for fluid obeying the energy conditions, having subluminal sound velocity and radial stability \cite{alho22}.\\


\section{Discussion}

We know that white dwarf is not supported by fluid pressure but instead by electron pressure and similarly neutron star by the Yang-Mills gauge fields. In the limiting case of the Buchdahl bound it is conceivable that distribution may attain free particles state of the Vlasov-like kinetic matter consisting of free particles in angular motion \footnote{However the total angular momentum remains zero retaining spherical symmetry of the distribution.} interacting only through gravity and the equilibrium is then governed by the Virial theorem, kinetic energy being equal to half potential energy. The Virial equilibrium state defines the limiting case of compactness which is attained only for the Buchdahl star. In general the compactness would be given by the parameter $\gamma \leq 1/2$. When $\gamma<1/2$, matter distribution has not yet attained the fully free particle kinetic state of the Vlasov distribution. \\

We would once again like to emphasise the fact that BS is physically accessible real astrophysical object. It could in fact be as real as black hole so long as there is no direct observational evidence for existence of horizon . What has been observed is a compact object \cite{bh-dis} not necessarily a BH but a compact object which could very well be BS. It therefore offers without any apology all that one was asking of the membrane or stretched horizon. All BH properties including thermodynamics could indeed be explored for BS. A couple of them have already been investigated. First, like BH \cite{dad-nar97}  non-extremal charged BS cannot be converted into an extremal one by adiabatic test particle accretion \cite{dad22a}. Second, a BH can though be over-extremalized under linear order accretion but the result is always overturned when second order perturbations are included \cite{sor-wal}; i.e., the weak cosmic censorship conjecture is always respected, the same has been shown to be true for BS as well \cite{dad22b}. Further it is an astrophysical object which is almost like BH, it could therefore serve as an excellent candidate for BH mimicker without any exotic and unusual properties. \\ 

Another remarkable property of BS is that its extremal limit is over extremal for BH. That is, BS could hold much more charge compared to BH because extremal bound for the former is $Q^2/M^2 = 9/8 >1$ which is over-extremal for BH. This gives rise to an interesting possibility, take a BS with $1<Q^2/M^2<9/8$ and let neutral matter accrete on it and thereby $Q^2/M^2$ could reduce down to $1$. If it does not trigger further collapse, it may end up in BS with $Q^2/M^2=1$. On the other hand if it does, it may lead to formation of an extremal BH. This is how extremality for BH which cannot be reached from the bottom, could now perhaps be attained from the top through BS. The important question to probe is whether an extremal BH could actually be formed this way? If it is indeed the case, BS could perhaps serve as precursor to BH in general and for extremal one in particular. However BS is the most compact object without horizon and it would therefore be an immediate state to BH which could perhaps turn into black hole on further accretion. \\

In this discourse on the Buchdahl star and the compactness bound, the following salient points that emerge are: (a) the role of gravitational energy in determining the bound, (b) the Virial theorem could govern the BS equilibrium and (c) all this could be done by using only the unique exterior metric without reference to interior at all. All these features are apriori unexpected and are therefore novel and insightful. The compactness of an object should naturally be determined by internal structure of fluid, its  equation of state, etc and above all the binding/internal energy. What we have shown is that measure of internal energy could be gauged by gravitational energy which could be computed from the exterior metric. This is a novel and illuminating realization that how compact a star can get to is governed by gravitational energy it has gained while collapsing from infinity to the radius it has now attained. Further BS defining condition is in fact the expression of the Virial equilibrium for the Vlasov like kinetic distribution. It is remarkable that all this could be done by appealing only to the unique exterior solution without reference to interior at all. In contrast there is no unique solution for interior except of course for constant density fluid, and that makes computation of internal energy from the interior solution formidably difficult. \\

Once again we would like to emphasize that like BH, BS could be a real astrophysical object which is very close to BH, and hence provides a rich astrophysical avenue which should be probed as BH mimicker and otherwise in various physical and astrophysical settings. In view of what we have argued all through, it would not be unreasonable to conjecture that in the limiting state of compactness interior distribution does attain the Vlasov-like kinetic state which is then governed by the Virial equilibrium condition. This emerges as a clear prediction.   \\

Further the important question to probe is what happens when matter accretes on BS, does it continue to retain its BS character or collapses down to become a BH? If the above picture of the Vlasov kinetic matter with its interior maintained by the Virial equilibrium, remains true, the accreting  matter would also have to have the same state for it to continue being BS. Else it would collapse to BH. For that accreting matter has to undergo a phase transition so as to accord to the Vlasov kinetic state. What would be required to effect such a transition? The same questions could be asked for accretion onto BH as well. These are very interesting and important issues with far reaching consequences for the compact objects (BH and BS) physics and astrophysics. This is what is going to engage us for quite some time to come.  \\


{\it Acknowledgements:} It is a pleasure to warmly thank Hakan Andreasson and Luciano Rezzolla for very helpful and engaging discussions that have considerably influenced the shape this manuscript has now taken. To the former I am particularly indebted for insights on the Einstein-Vlasov solutions and to the latter for clarifying many issues related to compact objects in general. \\


\begin{thebibliography}{99}
\bibitem{bh-dis} B.P. Abbott et al. (Virgo and LIGO Scientific Collaborations), Phys. Rev. Lett. {\bf 116}, 241102 (2016); arXiv:1602.03840 [gr-qc].
\bibitem{bh-dis1} S. Gillessen, P.Plewa, F. Eisenhauer, R. Sari, I. Waisberg, M. Habibi, O. Pfuhl, E. George, J. Dexter, S.vonFellenberg, T. Ott, R. Genzel, Ap. J. {\bf 837}, 30 (2017): arxiv:1611.09144. 
\bibitem{bh-dis2} A. Ghez, S. Salim, S. Hornstein, A. Tanner, J. Lu, M. Morris, E. Becklin, G. Duchene, Ap. J. {\bf 620}, 744 (2005); arxiv:astro-ph/0306130.
\bibitem{bh-sha} K. Akiyama et al. (Event Horizon Telescope Collaboration), Astrophys. J. {\bf 875}, L1 (2019); arXiv:1906.11238 [gr-qc].
\bibitem{thorne} R. Price, K. Thorne, Sci. Am. {\bf 258}, 69 (1988).
\bibitem{maxforce} N. Dadhich, Phys. Rev. {\bf D105}, 064044 (2022); arxiv:2201.10381.
\bibitem{buch} H. A. Buchdahl, Phys. Rev. {\bf 116}, 1027 (1959).
\bibitem{MTW} C. Misner, K. Thorne, J. Wheeler,  {\it Gravitation} ( W. H. Freeman, San Francisco, 1973). 
\bibitem{har-tho} J. Hartle, K. Thorne, Ap. J. {\bf 153}, 807 (1968).
\bibitem{dmk10} N. Dadhich, A. Molina, A. Khugaev, Phys. Rev. {\bf D81}, 104026 (2010), arxiv:1001.3922.
\bibitem{andrea08} H. Andreasson, J. Diff. Equations {\bf 245}, 2243 (2008).
\bibitem{kara-stal08} P. Karageorgis, J. Stalker, Class. Quant. Grav. {\bf 25} (2008) 195021.
\bibitem{mak-stuch} M. K. mak, P. N. Dobson, Jr, T. Harko, Mod, Phys. Lett. {\bf A15}, 2153 (2000);
T. Harko, M. K. Mak, J. Math. Phys. {\bf 41}, 4752 (2000); H. Andreasson, C. G. Boehmer, A. Mussa, Class. Quant. Grav.
{\bf 29}, 095012 (2012), arxiv:1201.5725; Z. Stuchlik, Acta. Phys. Slov. {\bf 50}, 219 (2000), arxiv:0803.2530.
\bibitem{andrea09} H. Andreasson, Commun. Math. Phys. {\bf 288}, 715 (2009).
\bibitem{ger-maa} C. Germani, R. Maartens, Phys. Rev. {\bf D64}, 124010 (2001), arxiv:hep-th/0107011.
\bibitem{gar-ure} M. A. Garca-Aspeitia, L. A. Urea-Lopez, Class. Quant. Grav. {\bf 32}, 025014 (2015), arxiv:1405.3932.
\bibitem{gos} R. Goswami, S. D. Maharaj, A. N. Nzioki, Phys. Rev. {\bf D92}, 064002 (2015), arxiv:1506.04043; J. Ponce de Leon, N. Cruz, Gen. Rel. Grav. {\bf 32}, 1207 (2000), arxiv:gr-qc/0207050; C. A. D. Zarro, Gen Re. Grav. {\bf 41}, 453 (2009).
\bibitem{wri} M. Wright, Gen. Rel. Grav. {\bf48}, 93 (2016), arxiv:1507.05560.
\bibitem{sum-dad16} S. Chakraborty, N. Dadhich, Phys. Rev. {D95}, 064059 (2017), arxiv:1606.01330.
\bibitem{Feng18} W. X. Feng, C. Q. Geng, L. W. Luo, Chin. Phys. {\bf C 43}, 083007 (2019): arxiv:1810.06753.
\bibitem{bar} D. Barraco, V. H. Hamity, Phys. Rev. {\bf D65}, 124028 (2002); A. Fujisawa, H. Saida, C. M. Yoo, Y. Nambu,
Class. Quant. Grav. {\bf 32}, 215028 92015), arxiv:1503.01517.
\bibitem{alho22} A. Alho, J. Natario, P. Pani, G. Raposo, Phys. Rev. {\bf D106}, 044025 (2022); arxiv:2202.00043.
\bibitem{roth} A. Giuliani, T. Rothman, Gen. Relativ. Grav. {\bf 40}, 1427 (2008).
\bibitem{mak-dob-har} M. K. Mak, P. N. Dobson Jr., T. Harko, Euro. Phys. Lett. {\bf 56}, 762 (2001).
\bibitem{boe-har} C. G. Boehmer, T. Harko, Gen. Relativ. Grav. {\bf 39}, 757 (2007).
\bibitem{and-ekl-rei} H. Andreasson, M. Eklund, G. Rein, Class. Quant. Grav {\bf 26}, 145003 (2009).
\bibitem{lemos14} J. P. S. Lemos, V. T. Zanchin, Class. Quant. Grav. {\bf 32}, 135009 (2015).
\bibitem{coop} F. I. Cooperstock, V. de la Cruz, Gen. Relativ. Grav. {\bf 9}, 835 (1978).
\bibitem{florides} P. S. Florides, J. Phys. {\bf A16}, 1419  (1983).
\bibitem{guil} B. S. Guilfoyle, Gen. Relativ. Grav. {\bf 31} (1999) 1645.
\bibitem{lemos10} J. P. S. Lemos, V. T. Zanchin, Phys. Rev. {\bf D81}, 124016 (2010).
\bibitem{lemos14a} J. D. V. Arbanil, J. P. S. Lemos, V. T. Zanchin, Phys. Rev. {\bf 89}, 104054 (2014).
\bibitem{bro-yor} J. D. Brown, J. W. York, Phys. Rev. {\bf D47}, 1407 (1993), arxiv:gr-qc/9209012.
\bibitem{dad97} N. Dadhich, Current Science {\bf 76}, 831 (1999), arxiv:gr-qc/9705037.
\bibitem{dad12} N. Dadhich, Current Science {\bf 109}, 260 (2015); arxiv:1206.0635.
\bibitem{ADM} R. Arnowitt, S. Deser, C. W. Misner, in Gravitation: an introduction to current research, ed. L. Witten (Wiley, New York, 1962); gr-qc/0405109.
\bibitem{maz-mot} P. Mazur, E. Mottola, Class. Quant. Grav. {\bf 32}, 215024 (2015), arxiv:1501.03806.
\bibitem{liu-yau} C-C. M. Liu, S-T. Yau, Phys. Rev. Lett. {\bf 23}, 231102 (2003).
\bibitem{dad19} N. Dadhich, JCAP {\bf 04}, 035 (2020).
\bibitem{sum-dad22} S. Chakraborty, N. Dadhich, Euro. Phys. J. {\bf C83}, 677 (2023); arxiv:2204.10734. 
\bibitem{Tiw17} A. Tiwari, S. Maharaj, R. Narain, Class. Quant. Grav. {\bf 34}, 155009 (2017).
\bibitem{And11} H. Andreasson, Living Rev. Rel. {\bf 14:4} (2011); arxiv:1107.1367.
\bibitem{And-Rei06} H. Andreasson, G. Rein, Class. Quant. Grav. {\bf 24}, 1809 (2007); arxiv:gr-qc/0611053.
\bibitem{Ame-And-Log19} E. Ames, H. Andreasson, A. Logg, Phys. Rev. {\bf D99}, 024012 (2019)' arxiv:1803.11224.
\bibitem{Ame-And-Log16} E. Ames, H. Andreasson, A. Logg, Class. Quant. Grav. {\bf 33}, 155008 (2016); arxiv:1603.05404.
\bibitem{dad-nar97} K. Narayan, N. Dadhich, Phys. Lett. {\bf A231}, 335 (1997).
\bibitem{dad22a} S. Shaymatov, N. Dadhich, arxiv:2209.02560.
\bibitem{sor-wal} J. Sorce, R. Wald, Phys. Rev. {\bf D96}, 104014 (2017); arxiv:1707.05862.
\bibitem{dad22b} S. Shaymatov, N. Dadhich, JCAP {\bf 06}, 010 (2023); arxiv:2205.01350. 

\end{thebibliography}
\end{document}